\documentclass[doublespacing]{elsart}

\usepackage{epsfig}

\usepackage{amssymb}

\begin{document}

\title{Neutron channelling in a magnetic tube trap}

\author{V.G. Baryshevsky}
\address{Research Institute for Nuclear Problems, 11 Bobryiskaya
str., 220050, Minsk, Belarus}

\maketitle

\begin{abstract}
Neutron channelling in a magnetic tube trap is discussed.
\end{abstract}

Quantum effects at motion of ultracold neutrons in a magnetic trap
are actively studied now.
In particular, ''magnetic tube'' traps \cite{1} are used for
investigation of such effects.
According to \cite{1}, the periodic set of magnetic traps can be
developed.
Traps in the form of ''magnetic tube'' are used for focusing of
atomic beams and described in details in \cite{2} on the basis of
the classic motion equations for a particle, possessing the
magnetic moment, in a magnetic field.
But quantum effects, which appear at neutron motion in the
magnetic trap, can not be described by classic motion equations.
The typical radius of the magnetic trap, which is being designed,
is $\rho \sim 10^{-4}$ cm \cite{1}.
 Thus, due to large wavelength ($\lambda \sim 10^{-5} \div 10^{-6}$ cm), motion of the ultracold neutron in the magnetic trap
 in the direction, transversal to the axis of the magnetic
tube, appears quantized.

Quantum effects were studied for neutron motion in a gravitational
trap \cite{3}.
Neutron motion in such a trap is one-dimensional.

But in contrast to \cite{3}, transversal motion of neutrons in the
''magnetic tube'' trap with respect to it axes is two-dimensional.
As noted below, neutron motion in such a trap is determined by
spin-independent effective potential energy, which is proportional
to the magnetic field strength square.

Thus, let us consider motion of the ultracold neutrons in the
tube-shaped magnetic trap.
The Schr{\"{o}}dinger equation, which describes the ultracold
neutron in such a trap, is as follows:
\begin{equation}
\left\{ -\frac{\hbar^2}{2 M} \Delta_r - \mu \vec{\sigma}
\vec{B}(\vec{r}) \right\} \Psi(\vec{r})= E \Psi(\vec{r})
 \label{Schrodinger}
\end{equation}
where $M$ is the neutron mass, $\mu$ is the neutron magnetic
moment, $\vec{r}$ is the coordinate of the neutron, $\vec{B}$ is
the magnetic induction in the point $\vec{r}$,
$\vec{\sigma}=(\sigma_x,\sigma_y,\sigma_z)$ are the Pauli
matrices.

Suppose the axes $z$ to be directed along the tube axes.
Neutron motion along the tube axes is quasi-classic.
Tube length (dimension in direction $z$) significantly exceeds its
radius.
Therefore, $\vec{B}(\vec{r})$ inside the tube can be considered as
independent on $z$, i.e
$\vec{B}(\vec{r})=\vec{B}(x,y)=\vec{B}(\rho,\varphi)$, where
$\rho$ and $\varphi$ are the radius and azimuth angle in the
cylindrical coordinate frame.

The equation (\ref{Schrodinger}) in the cylindrical coordinate
frame looks like as follows:
\begin{equation}
\left\{-\frac{\hbar^2}{2 M} \left[ \frac{1}{\rho}
\frac{\partial}{\partial \rho} \rho \frac{\partial}{\partial \rho}
+ \frac{\partial^2}{\partial z^2} + \frac{1}{\rho^2}
\frac{\partial^2}{\partial \varphi^2} \right] - \mu \vec{\sigma}
\vec{B}(\rho,\varphi) \right\} \Psi= E \Psi
 \label{Schrodinger1}
\end{equation}
Particle motion along $z$ is free and is described by the quantum
number $k_z$.
Therefore, the wave function of a neutron in the tube can be
expressed as:
\begin{equation}
\Psi(\vec{r})=\Phi(\rho) e^{i k_z z} = \sum_{m=-\infty}^{\infty}
\Phi_m (\rho) e^{i m \varphi} e^{i k_z z} \label{wavefunction}
\end{equation}
Therefore, to find the spectrum of neutron states in the tube the
following equation should be solved:
\begin{equation}
\left[ \frac{\partial^2}{\partial \rho^2} + \frac{1}{\rho}
\frac{\partial}{\partial \rho} - \frac{m^2}{\rho^2 } +
\varkappa^2\right] \Phi_m + \frac{2 M \mu}{\hbar^2}
\sum_{m^{\prime}} \vec{\sigma} \vec{B}_{m m^{\prime}} (\rho)
\Phi_{m^{\prime}} (\rho) =0 \label{phim}
\end{equation}
where $\varkappa^2 = k^2-k_z^2$, $k^2 = \frac{2 M E}{\hbar^2}$,
$\vec{B}_{m m^{\prime}} (\rho) = \frac{1}{2 \pi} \int e^{i
(m^{\prime}-m) \varphi} \vec{B} (\rho, \varphi) d \varphi$

Let us find the spectrum of neutron states in the trap in the
first order of perturbation theory.
In this case the diagonal matrix element of the interaction
energy, i.e. the average value of the magnetic field strength
$\vec{B}$ in the trap, determines the correction, which defines
spectrum.
But the averaged value of $\vec{B}$ in such a trap is zero.
Therefore, the nonzero contribution to the spectrum of the neutron
energy levels in the tube trap appears in the second order of the
perturbation theory.

For further analysis it is convenient to transform differential
equation (\ref{Schrodinger1}) to the homogeneous integral
equation:
\begin{equation}
\Psi(\vec{r}) = \int G_0(\vec{r},\vec{r}^{\prime})
\hat{V}(\vec{r}^{\prime}) \Psi{\vec{r}^{\prime}} d^3 {r}^{\prime}
 \label{integral}
\end{equation}
where $G_0(\vec{r},\vec{r}^{\prime})= - \frac{M}{2 \pi \hbar^2}
\frac{e^{i k
|\vec{r}-\vec{r}^{\prime}|}}{|\vec{r}-\vec{r}^{\prime}|}$ is the
Green function of the equation (\ref{Schrodinger1}) and the
potential energy $\hat{V}=-\mu \vec{\sigma} \vec{B}(\vec{r})$.
Substitution of the expression for $\Psi(\vec{r})$
(\ref{integral}) to the right part of equation (\ref{integral})
gives:
\begin{equation}
\Psi(\vec{r}) = \int G_0(\vec{r},\vec{r}^{\prime})
\hat{V}(\vec{r}^{\prime}) G_0(\vec{r}^{\prime},\vec{r}^{\prime
\prime}) \hat{V}(\vec{r}^{\prime \prime}) \Psi(\vec{r}^{\prime
\prime}) d^3 {r}^{\prime} d^3 {r}^{\prime \prime}
 \label{integral1}
\end{equation}

This equation is equivalent to the following expression

\begin{equation}
\left\{-\frac{\hbar^2}{2 M} \Delta_r - E \right\} \Psi(\vec{r}) +
\int \hat{V}(\vec{r}) G_0(\vec{r},\vec{r}^{\prime})
\hat{V}(\vec{r}^{\prime})  \Psi(\vec{r}^{\prime}) d^3 {r}^{\prime}
=0
 \label{Schrodinger2}
\end{equation}

Motion of the neutron along $z$ is quasiclassic, therefore the
quasiclassic expression for the Green function can be used:
\begin{equation}
G_0 (\vec{r},\vec{r}^{\prime}) = -i \frac{M}{\hbar^2 k} \delta
(\vec{\rho}^{\prime}-\vec{\rho}) \theta (z^{\prime} -z) e^{i k
(z^{\prime}-z)}
 \label{Green1}
\end{equation}
where $\theta (z^{\prime} -z)$ is the Heaviside unit function
($\theta (z^{\prime} -z)=1$, when $z^{\prime}>z$ and $\theta
(z^{\prime} -z)=0$, when $z^{\prime}<z$).

With the help of (\ref{Green1}) the expression
(\ref{Schrodinger2}) converts to
\begin{equation}
\left\{-\frac{\hbar^2}{2 M} \Delta_r - E \right\} \Psi(\vec{r}) -
i \frac{M}{\hbar^2 k} \hat{V}(\vec{\rho}) \hat{V}(\vec{\rho})
\int_z^{\infty} e^{-ikz} e^{ikz^\prime}
\Psi(\vec{\rho},z^{\prime}) d z^{\prime} =0
 \label{Schrodinger3}
\end{equation}

It should be mentioned that $\hat{V}(\vec{\rho})
\hat{V}(\vec{\rho})=\mu^2 (\vec{\sigma} \vec{B})(\vec{\sigma}
\vec{B}) = \mu^2 \vec{B}^2 (\vec{\rho})$.

With the help  (\ref{wavefunction}) the equation
(\ref{Schrodinger3}) can be expressed as
\begin{equation}
\left\{-\frac{\hbar^2}{2 M} \Delta_{\rho} - \frac{\hbar^2
\varkappa^2}{2 M} \right\} \Phi(\vec{\rho}) + V_{eff}(\vec{\rho})
\Phi(\vec{\rho}) =0
 \label{Schrodinger4}
\end{equation}
where
\begin{equation}
V_{eff}(\vec{\rho}) = \frac{M \mu^2 B^2 (\vec{\rho})}{\hbar^2 k
(k+k_z)}
\label{Veff}
\end{equation}
is the effective potential energy.

According to (\ref{Veff}) neutron motion in the magnetic tube
trap, which is formed by the alternating magnetic field, is
determined by the squared magnetic field strength and does not
depend on the neutron spin direction.

The equation (\ref{Schrodinger4}) allows to find the
eigenfunctions and the spectrum of eigenvalues of neutron energy
in the trap.

In the cylindrical coordinate frame the equation
(\ref{Schrodinger4}) reads as follows:
\begin{equation}
\hspace{-1.5 cm} \left( \frac{\partial^2}{\partial \rho^2} +
\frac{1}{\rho} \frac{\partial}{\partial \rho} - \frac{M^2}{\rho^2}
+ \varkappa^2 \right) \Phi_m (\rho) - \frac{2 M}{\hbar^2}
\sum_{m^{\prime}} \langle m | \hat{V}_{eff} (\vec{\rho}) |
m^{\prime} \rangle \Phi_{ m^{\prime}} (\vec{\rho}) = 0
\label{Schrodinger5}
\end{equation}
here $\Phi(\vec{\rho}) = \sum_m \Phi_m (\vec{\rho})
e^{im\varphi}$, $\langle m | \hat{V}_{eff} (\vec{\rho}) |
m^{\prime} \rangle = \frac{1}{2 \pi} \int_0^{2 \pi} V_{eff}
(\vec{\rho},\varphi) e^{-i (m-m^{\prime}) \varphi} d \varphi$.

Let us keep in the sum in (\ref{Schrodinger5}) only the term with
$m^{\prime} = m$,
 which describes the effective potential energy
 $V_{eff} (\vec{\rho}, \varphi)$ averaged over the azimuth angle
 $\varphi$.
(More detailed analysis of neutrons in the trap could require
consideration of additional  terms in the sum with $m^{\prime} \ne
m$.)
Then (\ref{Schrodinger5}) looks like as follows:
\begin{equation}
 \left( \frac{\partial^2}{\partial \rho^2} +
\frac{1}{\rho} \frac{\partial}{\partial \rho} - \frac{M^2}{\rho^2}
+ \varkappa^2 \right) \Phi_m (\rho) - \frac{2 M}{\hbar^2} \langle
m | \hat{V}_{eff} (\vec{\rho}) | m \rangle \Phi_{ m} (\vec{\rho})
= 0
\label{Schrodinger6}
\end{equation}
According to \cite{?} magnetic field in the tube trap  linearly
decreases to the trap center, i.e. $\vec{B} (\vec{\rho}) \sim
\rho$. Therefore, $B^2 (\vec{\rho}) \sim \rho^2$ and $\langle m |
\hat{V}_{eff} (\vec{\rho}) | m \rangle \sim \rho^2$.
Therefore, equation (\ref{Schrodinger6}) can be rewritten as:
\begin{equation}
\left( \frac{\partial^2}{\partial \rho^2} + \frac{1}{\rho}
\frac{\partial}{\partial \rho} - \frac{M^2}{\rho^2} + \varkappa^2
\right) \Phi_m (\rho) - A \rho^2 \Phi_{ m} (\vec{\rho}) = 0
\label{Schrodinger7}
\end{equation}
where $A$ is determined by
\begin{equation}
A \rho^2 = \frac{2 M}{\hbar^2} \langle m | \hat{V}_{eff}
(\vec{\rho}) | m \rangle =
\frac{2 M \mu^2 \langle m | B^2 (\vec{\rho}) | m \rangle}{\hbar^4
k(k+k_z)}
\label{A}
\end{equation}

The eigenvalues and eigenfunctions of the obtained equation
(\ref{Schrodinger7}) are the those of the cylindrically symmetric
oscillator \cite{4}.
Introducing new variable $\xi = \sqrt{A} \rho^2$ one can get from
(\ref{Schrodinger7}) the following equation:
\begin{equation}
\xi \frac{\partial^2 \Phi_m}{\partial \xi^2} + \frac{\partial
\Phi}{\partial \xi} + (-\frac{\xi}{4} + \beta -\frac{m^2}{4
\xi})\Phi_{m}  = 0
\label{Schrodinger8}
\end{equation}
where $\beta=\frac{\varkappa^2}{4 \sqrt{A}}$.

Solution of this equation is as follows \cite{?}:
\begin{equation}
\Phi_{m} (\xi)   = e^{-\frac{\xi}{2}} \xi^{\frac{|m|}{2}} W(\xi)
\label{solution}
\end{equation}
where $W(\xi)$ is the confluent hypergeometric function
\begin{equation}
W(\xi)= F \left\{ -( \beta - \frac{|m|+1}{2} ), |m|+1, \xi
\right\}.
\label{W}
\end{equation}

To make the  wavefunction $\Phi_m$ finite everywhere, the
parameter $\beta - \frac{|m|+1}{2}$ should be integer nonnegative
number, i.e.
\begin{equation}
\beta - \frac{|m|+1}{2}= n_{\rho}, \textrm{~where~} n_{\rho} \ge
0.
\label{condition}
\end{equation}
This condition (\ref{condition}) determined spectrum of the
transversal motion for the neutron in the magnetic trap.

To avoid misunderstanding, I'd like to emphasize that I consider
spectrum of the energy levels, which are lower then the height of
the effective potential barrier.

Substitution $\beta=\frac{\varkappa^2}{4 \sqrt{A}}$ in the
explicit form to (\ref{condition}) provides to get:
\begin{equation}
\frac{\varkappa^2}{4 \sqrt{A}} - \frac{|m|+1}{2}= n_{\rho},
\textrm{~where~} n_{\rho} \ge 0,
\label{condition1}
\end{equation}
therefore
\begin{equation}
E_{trans} (n_{\rho}, m) = \frac{\hbar^2 \varkappa^2 (n_{\rho},
m)}{2 M} = \frac{\hbar^2}{2 M} 4 \sqrt{A} (n_{\rho} +
\frac{|m|+1}{2}) \label{Etrans}
\end{equation}

Let us consider the particular example. The averaged field can be
expressed as
\begin{equation}
\langle m| B^2 (\vec{\rho}) | m \rangle = \frac{\langle B^2_{max}
 \rangle}{\rho^2_{max}} \rho^2
\end{equation}
where $\rho_{max}$ is the radius of the magnetic tube, $B^2_{max}$
is the square averaged magnetic field at $\rho=\rho_{max}$.
In this case the coefficient $A$ is
\begin{equation}
A= \frac{2 M^2}{\hbar^2} \frac{\mu^2 \langle B^2_{max}\rangle}{k
(k+k_z) \rho^2_{max}}
 \label{A1}
\end{equation}

 The frequency of the  transverse oscillations of the neutron in the ground state  in the magnetic tube
 \begin{equation}
\Omega_{trans}= \frac{\hbar \sqrt{A}}{M} = \sqrt{2} \frac{\mu
\sqrt{\langle B^2_{max} \rangle}}{\hbar \sqrt{k (k+k_z)}
\rho_{max}}
 \label{frequency}
\end{equation}

For the ultracold neutrons with $\lambda \approx 10^{-6} \div
10^{-5}$ cm in the tube with $\rho_{max} \sim 10^{-4}$ cm the
product $\sqrt{k (k+k_z)} \rho_{max} \approx k \rho_{max} \gg 1$.
Therefore, for the magnetic field $\sqrt{\langle B^2_{max}
\rangle} = 10^4$ gauss the frequency $\Omega_{trans} \approx 10^4
\div 10^5$ s$^{-1}$.
Suppose that during the time $\tau$ the neutron moves with the
speed $10^2$ cm/s in the tube of 1 cm length (time of neutron
flight in the trap is about $\tau \sim 10^{-2}$ s), then  the
parameter $\Omega_{trans} \tau \gg 1$, i.e. neutron executes many
oscillations in the trap.

Thus, neutron motion in the magnetic tube is described by the
energy of the transversal motion $E_{trans}$ and the energy of
longitudinal motion $E_{long}=\frac{\hbar^2 k_z^2}{2 M}$.
The sum of these energies is equal to the neutron energy before it
flies into the trap $E=\frac{\hbar^2 k^2}{2 M}$:
\begin{equation}
E=E_{trans}+E_{long}. \label{E}
\end{equation}
The equation (\ref{E}) can be used to find $k_{\|}$ for the
neutron in the certain state of transversal motion.

When the energy of transversal motion is comparable or exceeds the
height of the effective potential barrier, the above expressions
for the wavefunctions and $E_{trans}$ cannot be applied.

%
The typical angle of neutron incidence on the magnetic tube trap
providing the energy of transversal motion to be less than the
height of the effective potential barrier can be expressed as:
\begin{equation}
\vartheta_{cr} \sim \sqrt{\frac{V_{eff}}{E}} \approx
\frac{\Omega_{L}}{E/\hbar}. \label{angle}
\end{equation}
where $\Omega_{L}$ is the frequency of Larmor precession of
neutron spin in the field $\sqrt{\langle B^2_{max} \rangle}$.
The expression (\ref{angle}) was obtained considering $k \approx
k_z$ (because $k_z$ converges to $k$ with energy growth).

According to (\ref{angle}) the neutron moving with the speed 1 m/s
in the trap with the maximal field $B \approx 10^{4}$ gauss can be
captured to the trap with the potential $V_{eff}$ at the angle
$\vartheta < \vartheta_{cr} \approx 1$ rad.
When neutron speed is $v=10$ m/s, the angle $\vartheta_{cr}
\approx 10^{-2}$ rad.
Energy growth results in reducing of $\vartheta_{cr} \sim
\frac{1}{E} \sim \frac{1}{v^2}$

Neutron passing through the magnetic tube trap is similar to the
process of axial channelling of a charged particle in a crystal.
Study of the angular distribution for neutrons passing through the
trap provides to find spectrum of the neutron levels in the trap.

Actually, experiments for neutrons(atoms) passing through the
magnetic tube trap can be considered as experiments studying
neutron (atom) channelling in such a trap.
When the traps are periodically distributed in the plane $(x,y)$
the analogy becomes ever more evident.

It should be also mentioned that bent magnetic tube traps can be
used for focusing wide beams of ultra cold neutrons (similar
focusing of particles channelled in a crystal).

\section{Acknowledgements}

I wish to thank V.F. Ezhov, who initiated topic discussion.

\end{document}